\documentstyle[11pt,paspconf,epsf]{article}

\begin{document}

\title{Panel Review: Reconstruction Methods and
the Value of $\beta$}

\author{Jeffrey A.\ Willick}
\affil{Physics Department, Stanford University, 
    Stanford, CA 94305-4060}

\begin{abstract}
The closely related issues of (1) reconstructing density and
velocity fields from redshift survey and distance indicator data
and (2) determining cosmological parameters from such data sets
are discussed. I focus on possible explanations for
the discrepant values of $\beta\equiv\Omega^{0.6}/b$
resulting from analyses that employ various reconstruction methods.
Although no firm conclusions are reached, possible routes towards 
resolution of these discrepancies
are suggested by the discussion.

\end{abstract}


\section{Introduction}

The original charge to our panel, which consisted of
myself, Marc Davis, Avishai Dekel, and Ed Shaya, was
to discuss ``Reconstruction Methods.'' To
narrow the discussion, we decided in
advance to concentrate on the problem of deriving $\beta$
from comparison of redshift survey and distance indicator
data. This seemed appropriate given the panel membership:
all of us have worked on this
problem, and together we span the spectrum of results,
from low values of $\beta$ suggestive of a low
density ($\Omega_M\sim 0.2$) universe to high values
consistent with $\Omega_M=1.$ 

While it may appear that we have strayed from the theme
of ``reconstructions,'' 
$\beta$-determination is, in fact,
closely linked with how we reconstruct the underlying
velocity and density fields from the observable data.
I clarify this point in \S~2 below. In \S~3, I review
recent results for $\beta$ from peculiar
velocities and consider
possible explanations for their wide variance. 
I conclude in \S~5 with a brief 
look to the future of the subject.

\section{Reconstructions and $\beta$}

Our theoretical formulation of cosmic dynamics involves 
the mass overdensity field $\delta_\rho(\vec r)$
and the peculiar velocity field ${\vec v_p}(\vec r).$ Linear
gravitational instability theory tells us that
these fields obey the simple relation 
$\nabla\cdot {\vec v_p} = -f(\Omega_M)\delta_\rho,$
where $f(\Omega_M) \approx \Omega_M^{0.6}.$ 

Such continuous fields are abstract
mathematical constructs; reality is more complex.
Discrete entities---galaxies---sparsely populate 
the universe, and we measure their redshifts and estimate
their distances. {\em Reconstruction Methods\/} are the
numerical algorithms we employ to turn these real-world
data into representations of the underlying 
fields $\delta_\rho$ and ${\vec v_p}.$

Several obstacles
stand between the measurements and successful reconstruction
of the density and velocity fields. 
First, there is {\bf biasing:}
how is the galaxy overdensity $\delta_g$ related to
the desired $\delta_\rho?$ 
In the oft-used if oversimplistic
linear biasing paradigm $\delta_g=b\,\delta_\rho,$
and the linear velocity-density relation becomes
$\nabla\cdot {\vec v_p} = -\beta \delta_g.$ To the
extent linear biasing and linear dynamics hold,
only $\beta,$ not $\Omega_M,$ is measureable by
this method. 
Incorporating ``realistic'' models of biasing---and
it is by no means clear that such models exist yet---presents
another challenge to reconstruction methods.
Reconstruction
of ${\vec v_p}$ is if anything an even more
daunting task: we estimate only the radial component
of this field, at discrete, irregularly distributed
positions, with large errors that grow linearly with distance. 
 
We cannot
invoke the theoretical velocity-density relation,
and thus measure $\beta,$
until we have reconstructed the underlying fields. 
If our reconstructions of $\delta_g(\vec r)$
and/or ${\vec v_p}(\vec r)$ are flawed, we'll get
the wrong value of $\beta.$ And given the 
difficulties such reconstructions face, 
errors in $\beta$ are, perhaps, inevitable
at this early stage in our understanding.

\begin{figure}[th!]
\textwidth=4.25in
\plotone{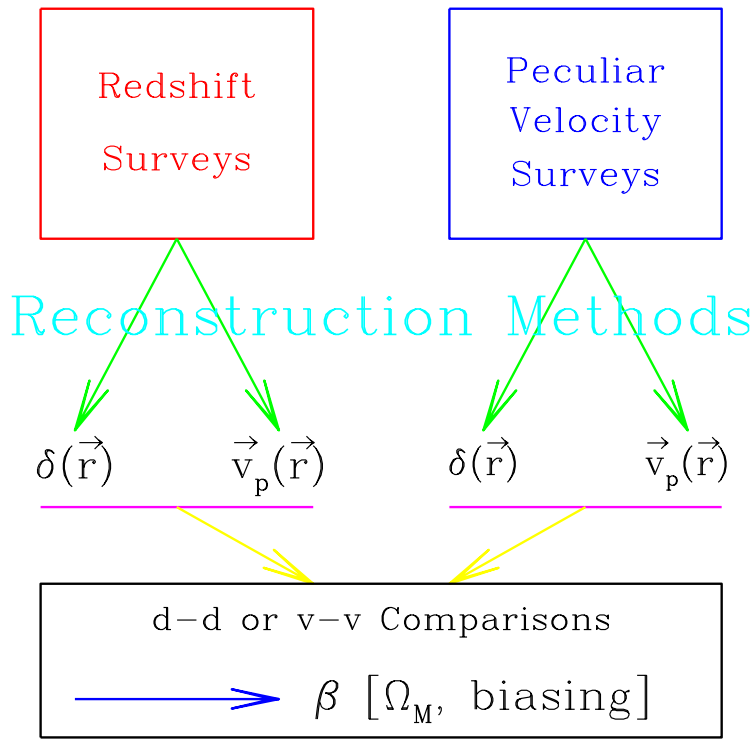}
\caption{A schematic representation of the relationship between
input data, reconstructions, and derivation of cosmological
parameters.}
\label{fig:recon}
\end{figure}

Figure~\ref{fig:recon} summarizes the relationship
between reconstruction methods and $\beta$-determination.
The box at the bottom concerns
the nature of the comparison by which $\beta$ is obtained.
If one reconstructs
$\vec v_p$ from the peculiar velocity data 
and compares its divergence 
with $\delta_g$ from redshift survey
data, one is doing 
a {\em density-density\/} (d-d) comparison.
Alternatively one can reconstruct 
${\vec v_p}(\vec r)$ from redshift survey data (for
an assumed value of $\beta$), using the
integral form of the linear-velocity density relation, and compare
with observed radial peculiar velocity estimates.
This is known as the {\em velocity-velocity\/} (v-v) comparison.
The distinctions between the two approaches are subtle but
important, as discussed below.

\textwidth=5.25in

\section{Discrepant $\beta$ Values, and Possible Explanations for Them}

It was not until the early 1990s that measuring $\beta$
by comparing peculiar velocity and redshift survey data
became a realistic goal. The advent of full-sky redshift
surveys---notably
those based on the IRAS point source catalog---and
large, homogeneous sets of
Tully-Fisher (TF) and related data were
the key developments. The earliest attempt 
was that of Dekel et al.\ (1993),
the so-called ``POTIRAS'' comparison. In this procedure
the velocity field is reconstructed using the POTENT
algorithm, and its divergence compared with the galaxy
density field from IRAS (a d-d comparison). 
Dekel et al.\ (1993) found
$\beta_I=1.29,$\footnote{Here
and below we apply the subscript $I$ to $\beta$ when the
galaxy density field is obtained from IRAS. Other redshift surveys
yield different overdensities and
thus different $\beta$s.}
a value widely taken
as support for the then-popular Einstein-de Sitter paradigm.
The POTIRAS analysis has since been redone with much improved
peculiar velocity data, with the result
$\beta_I=0.89 \pm\ 0.1$ (Sigad et al.\ 1998), still
consistent with a critical density universe unless IRAS galaxies
are strongly anti-biased. 

Between the first and second POTIRAS papers, a number
of studies, based on the v-v comparison, 
arrived at markedly lower values of $\beta_I.$
Shaya, Tully, \& Peebles (1995) estimated $\Omega_M=0.17\pm 0.1$
($\beta_I\approx 0.35$ if IRAS galaxies are unbiased) using
the Least Action Principle to predict peculiar velocities.
Willick et al.\ (1997) and Willick \& Strauss (1998)
used the VELMOD method to find $\beta_I=0.5\pm 0.06.$ 
The above results were obtained from TF data; an application
of similar methods to more accurate (but far more sparse)
supernova data yielded $\beta_I=0.4\pm 0.1$ (Riess et al.\ 1998).
(This list is not exhaustive, and apologies
are due to authors not cited; my aim is illustrative rather
than comprehensive.)

Thus, it appears that a somewhat bimodal distribution of
$\beta$ values has emerged. The d-d, POTIRAS method
produces $\beta_I$ close to unity; the v-v comparisons,
several based on the {\em same\/} redshift and velocity
samples as POTIRAS, yield $\beta_I\sim 0.5.$ Neither
the d-d nor the v-v comparison is inherently more valid;
both are firmly grounded in linear gravitational instability theory.
What, then, could be the cause of the discrepancies?

Neither the panelists nor the conference participants
arrived at a satistfactory answer to this question.
However, the following list of possible explanations
proved
to be fertile ground for discussion. The list
is accompanied by my own (biased)
commentary on the salience of each explanation. 
\medskip
\begin{itemize} 
\item {\bf Malmquist bias.} This famous statistical effect
has long been invoked as a cause of discrepant
conclusions in astronomy. In fact, it is now a non-issue.
Methods are now available for rendering
Malmquist and related
biases inconsequential; 
see Strauss \& Willick (1995) for a detailed discussion. 

\item {\bf Nonuniverality of the TF and other distance-indicator
relations.} Another non-issue.
It is easy to make the general argument---nonuniversal
distance indicators imply spurious peculiar velocities---but
there is little or no evidence implicating such
effects in the $\beta$ problem. 

\item {\bf Calibration Errors in Peculiar Velocity Datasets.} 
Such errors may indeed be present; new results from
the Shellflow project (see the paper by Courteau in these
proceedings) suggest that the widely used Mark III catalog
has across-the-sky calibration errors.
However, these errors affect mainly bulk flow estimates,
not the value of $\beta_I$ (see Willick \& Strauss 1998 for
a clear demonstration of this).

\item {\bf Non-trivial biasing.} It is now widely believed
that biasing is not only nonlinear, but stochastic
and nonlocal (see the panel review by Strauss in 
these proceedings). If so, 
it is perhaps unsurprising that different approaches
produce different results, as they may in fact be
measuring different things. 
This argument may well contain a kernel of truth,
but I am not convinced it is the whole story.
I was impressed by the work of Berlind, Narayanan, and
Weinberg on this subject (see their contributions
to the proceedings), which shows that for all but the
most contrived models of biasing,
the value of $\beta$ one obtains
is relatively insensitive to methodology.

\item {\bf The density-density versus the velocity-velocity comparison.}
This, I think, is the central issue. We may be asking
too much of our distance indicator data when we
use them to derive the full 3-D velocity field and
its derivatives, as is required for the d-d comparison. 
In the v-v comparison, by contrast, 
the distance indicator data is
used essentially in its raw form; only the redshift
survey data, which is intrinsically more accurate,
is subject to complex, model-dependent manipulation.
From this perspective, I would argue that the
low values, $\beta_I \approx 0.4$--$0.5,$ 
that have come out of 
the v-v analyses are more likely to be correct.
\end{itemize}

\section{A Quick Look to the Future}

If we learned anything from our panel discussion, it
was the usual lesson: better data will help. Fortunately,
some already exist, and more are on the way.
The Surface Brightness Fluctuation (SBF) data 
of Tonry and collaborators (see Tonry's paper in
these proceedings) are now available for comparison
with the IRAS redshift data; preliminary
results are reported by Blakeslee in these proceedings.
SBF distances are considerably more accurate
than TF distances and promise a much higher-resolution
look at the velocity field.  Adam Riess reported
new, and also very accurate, 
SN Ia distances for nearby galaxies.
The extant TF 
data are not likely to increase dramatically
in the short term, but will be recalibrated by
the Shellflow program, and perhaps the different
TF data sets (Mark III, SFI, Tully's Catalog) will be merged
into a larger, homogeneous catalog.  
In the longer term ($\sim 5$--10 years), they
will be supplanted by much larger and more uniform
TF data sets that will emerge from the wide-field
infrared surveys currently under way, 2MASS and
DENIS (see the contributions by Huchra and Mamon
in these proceedings).

I draw yet another
lesson from our panel discussion: we have
not yet fully developed the analytical methods
needed to deal with nonlinearities in the
universe, both dynamical and biasing-related.
In this regard I would encourage theorists
to address the problem of, how, given a
particular biasing model, do we predict
the peculiar velocity field from redshift survey
data, taking full account of nonlinear dynamics?
When reliable methods are developed for doing
this, and tested against N-body simulations,
we will be able to fully exploit the improved
peculiar velocity data sets of the future.

\acknowledgments
I thank my fellow panel members, Marc Davis, Avishai Dekel, and
Ed Shaya, for stimulating discussions.

\end{document}